\documentclass[runningheads]{llncs}
\usepackage{graphicx}
\usepackage{hyperref}

\usepackage{microtype}

\usepackage{latexsym}
\usepackage{setspace}
\usepackage[misc]{ifsym}
\usepackage{verbatim}
\usepackage{caption}
\usepackage{graphicx}
\usepackage{color}
\usepackage{multirow}
\usepackage{tabularx}
\makeatletter
\newcommand*{\addFileDependency}[1]{
  \typeout{(#1)}
  \@addtofilelist{#1}
  \IfFileExists{#1}{}{\typeout{No file #1.}}
}
\makeatother

\newcommand*{\myexternaldocument}[1]{%
    \externaldocument{#1}%
    \addFileDependency{#1.tex}%
    \addFileDependency{#1.aux}%
}

\usepackage{xr}
\myexternaldocument{supplementary}

\usepackage{times}
\usepackage{latexsym}
\usepackage{subcaption}

\usepackage{amsmath}
\usepackage{array}
\usepackage{algorithm}
\usepackage{algorithmic}

\newcolumntype{L}{>{\centering\arraybackslash}m{5cm}}
\newcounter{fnote}

\usepackage[T1]{fontenc}

\usepackage[utf8]{inputenc}

\usepackage{microtype}

\title{MWPRanker: An Expression Similarity Based Math Word
Problem Retriever}

\author{Mayank Goel  \and Venktesh V \Letter\footnotemark[1] \and
Vikram Goyal}

\tocauthor{Mayank ~ Goel , Venktesh ~ V ,  Vikram ~ Goyal}
\toctitle{Self-Supervised Paraphrase Quality Detection for Algebraic Word Problems}
\institute{Indraprastha Institute of Information Technology, Delhi \email{\{mayank.co19\}@nsut.ac.in,\{venkteshv,vikram\}@iiitd.ac.in}
}

\begin{document}
\maketitle
\begin{abstract}
    Math Word Problems (MWPs) in online assessments help test the ability of the learner to make critical inferences by interpreting the linguistic information in them. To test the mathematical reasoning capabilities of the learners, sometimes the problem is rephrased or the thematic setting of the original MWP is changed. Since manual identification of MWPs with similar problem models is cumbersome, we propose a tool in this work for MWP retrieval. We propose a hybrid approach to retrieve similar MWPs with the same problem model. In our work, the problem model refers to the sequence of operations to be performed to arrive at the solution. We demonstrate that our tool is useful for the mentioned tasks and better than semantic similarity-based approaches, which fail to capture the arithmetic and logical sequence of the MWPs. A demo of the tool can be found at \url{https://www.youtube.com/watch?v=gSQWP3chFIs}
\end{abstract}

\section{Introduction}
Math Word Problems (MWPs) are intriguing as they require one to decipher the problem model and operators from the given problem statement. Studies have shown that users lacking this ability often commit mistakes when presented with new problems \cite{mwp_lt}. It has been demonstrated that solving paraphrased versions of the original problem might aid in better learning to make critical inferences from varying linguistic information \cite{mwp_paraphrasing}. However, manual curation of such problems is cumbersome. Hence, we design a tool in this work to recommend problems with similar algebraic expressions as the input MWP.

\begin{figure}
    \centering
    \includegraphics[width=0.8\linewidth]{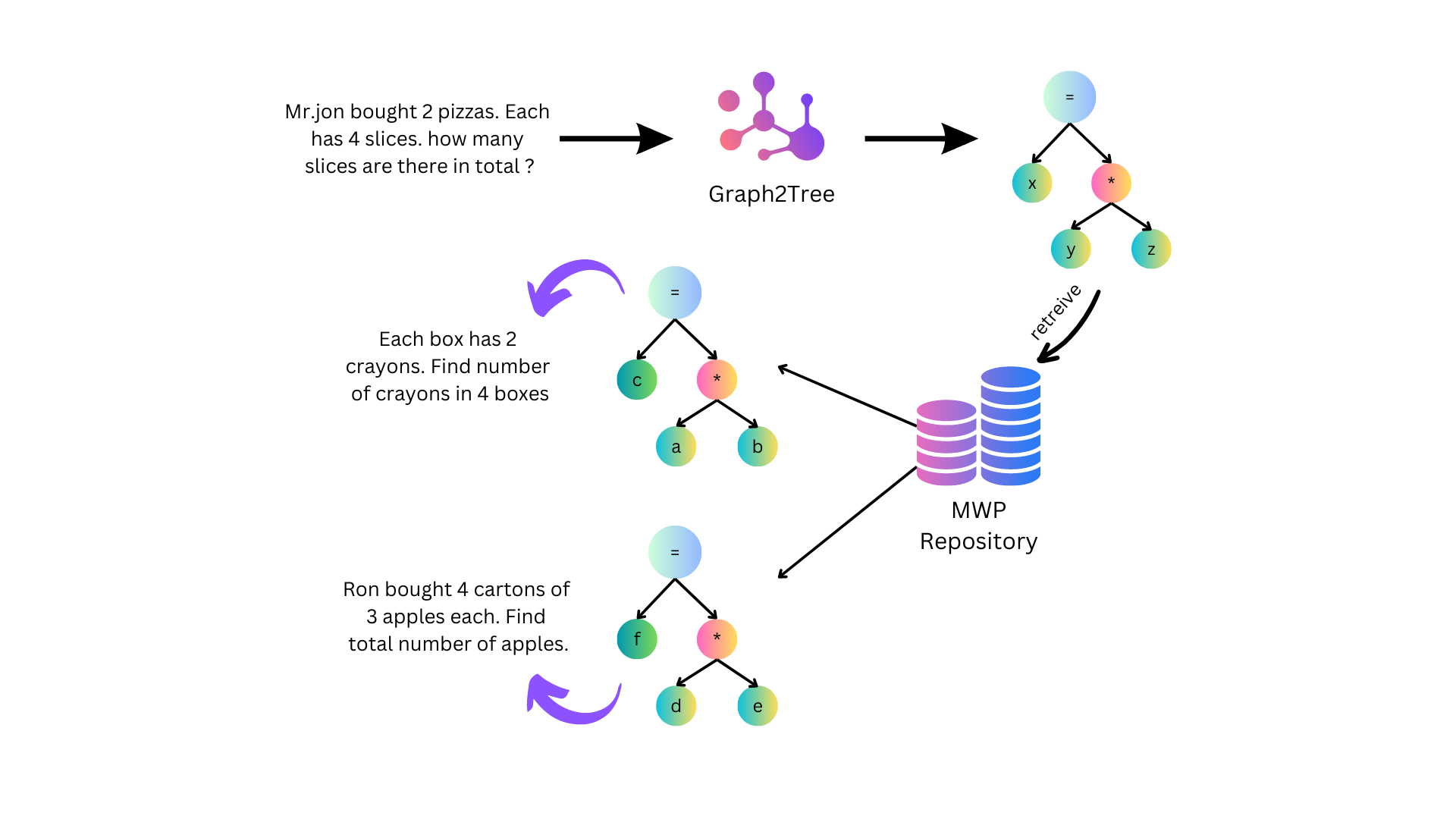}
    \caption{MWPRanker system pipeline}
    \label{fig:my_label}
\end{figure}

Numerous works have tackled the automated solving of Math Word Problems (MWPs) task \cite{mwp_solving1}, \cite{statistical_mwp_solver},\cite{contrastive_mwp_solver}. Recently, large language models (LLMs) have demonstrated multi-step reasoning ability to solve MWPs \cite{llm_multistep} among other tasks. However, the authors, in their work \cite{mwp_verifiers}, demonstrate that LLMs fail when problems contain certain linguistic variations. However, very few works have dealt with the retrieval of MWPs. Certain works like \textit{Recall and Learn} \cite{huang-etal-2021-recall-learn} leverage the task of retrieving analogous problems to solve MWPs using semantic similarity. However, they may wrongly recommend problems with different algebraic operations. For instance, ``John had 5 apples, and Mary had 6 oranges. Find the total number of fruits" would be considered similar to ``John had 5 apples, and Mary had twice as many oranges after selling 2 of them. Find the total number of fruits". Though these MWPs look similar, the second MWP requires additional multiplication and subtraction operation. 

An overview of the workflow of the proposed system is shown in Figure 1. The core contributions of our work are:
\begin{itemize}
    \item We propose a hybrid approach to retrieve similar Math Word Problems based on expression tree similarity. 
    \item The code and data can be found at  \url{https://github.com/goelm08/MWP-ranker}
    \end{itemize}

    \section{System Design}

    In this section, we describe the proposed system for identifying similar Math Word Problems for a given input. Given a corpus of MWPs $P= \{p_1,p_2....p_n\}$ and a new input MWP sequence $p_{new}={w_1,w_2...w_n}$, the goal is to recommend exact duplicate problems $p_{dup}$ based on expression similarity.

In our proposed pipeline, two problems could be similar if they are paraphrased versions of each other but evaluate the same algebraic expression.
We propose a hybrid pipeline \textit{MWPRanker} which is efficient for similar MWP retrieval. The pipeline consists of the following stages.

\begin{itemize}
    \item The input MWP $p_{new}$ is parsed into an algebraic expression $a_{new}$ using the neural expression generator \textit{Graph2Tree}. We derive an expression tree from the resulting expression.
    \item We devise a tree matching algorithm to match the resulting tree with other expression trees in the repository. The MWPs corresponding to matching expressions are returned to the user.  The expressions trees are derived and indexed for efficient retrieval at inference time. This is a one time activity as this is the repository our pipeline performs search on.
\end{itemize}

\subsection{Generating Expression from MWPs}
We employ Graph2Tree \cite{graph2tree} with minor variations.
Graph2Tree leverages the dependency graph of the input MWP with minor variations and translates it to an algebraic expression using a decoder model. First, the dependency parse of the input MWP is obtained using Stanford CoreNLP \cite{manning-etal-2014-stanford}. We identify the keyphrases in the sentence, such as noun phrases, and establish relationships between them. The relationships indicate important linkages and are created as separate nodes. This yields a heterogeneous graph. Then BiGraphSAGE is employed to compute graph-based contextualized embeddings. BiGraphSAGE is a variation of GraphSAGE \cite{graphsage}, including forward and reverse mode aggregation for computing node embeddings. Then a Bi-LSTM is employed as a decoder to generate the expression sequence leveraging the node embeddings from the graph. The decoder at inference time yields an algebraic expression $y_{exp}$ which is sent to the tree generator and matching module for retrieving similar MWPs. We adapt and modify the implementation of Graph2Tree in MWPToolkit \cite{lan2021mwptoolkit} with same hyperparameters.

\subsection{Tree Matching and Retrieval}
\begin{figure}
    \centering
    \includegraphics[width=0.7\linewidth]{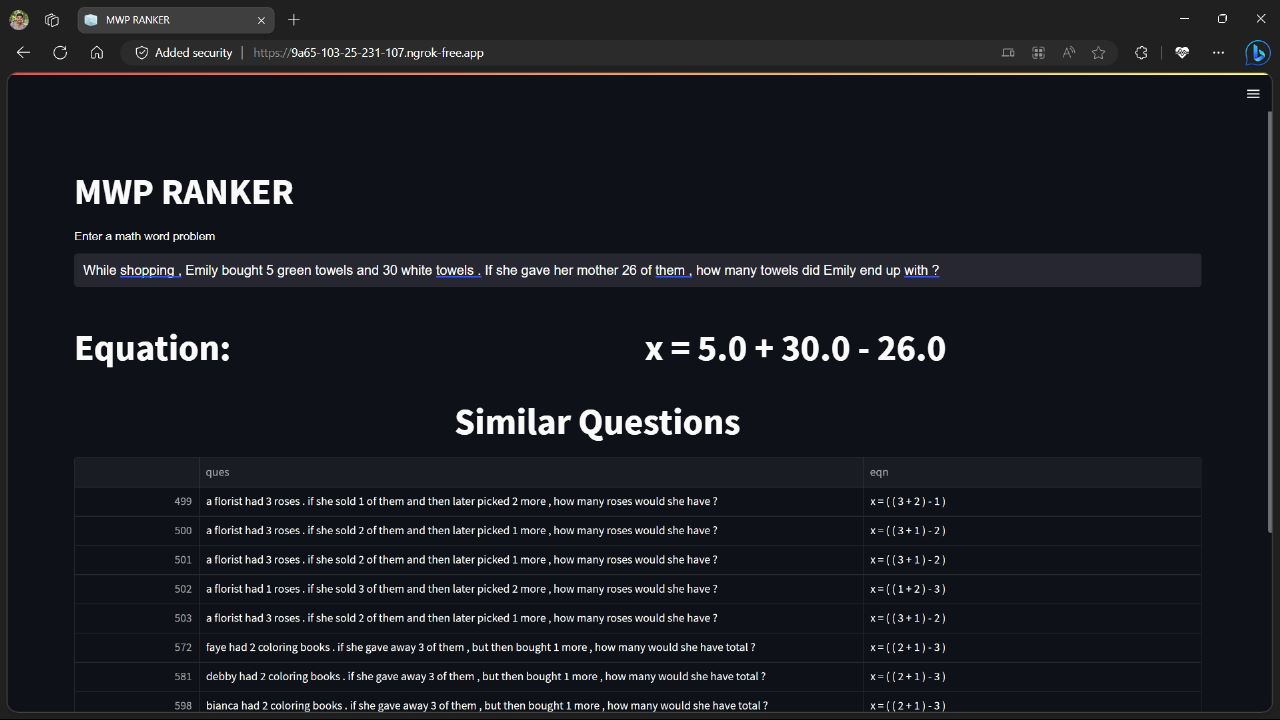}
    \caption{MScreenshot of MWPRanker}
    \label{fig2}
\end{figure}

We derive a postfix expression and convert it to an expression tree for clear operator precedence.
$t_{exp} = f_{tree}(y_{exp}) $. We replace numbers that represent variables with variable names. We replace constant values with the expression "<CONSTANT>"
The generated expression tree is compared with other expression trees $T_{exp}  = \{t_{exp}^1....t_{exp}^n\}$ in the MWP repository and the top-k problems are recommended to the user (Figure \ref{fig2}).

The matching algorithm works as follows:
\begin{itemize}
    \item The expression trees are matched pairwise through postorder traversal.
    \item If the node in a tree contains an operator, the other tree must contain the same operator in the same place.
    \item In the same way, variable nodes must match. If  a variable is encountered, the corresponding node in the other tree should also be a variable. When encountering a constant, we just check if the corresponding node in the other tree also contains the same placeholder.
     When a match is encountered, the corresponding natural language form of the MWP from the repository is returned.
\end{itemize}

\section{Demonstration}
We train the models using PyTorch. The models are served through a Flask API as backend, and UI is designed using Streamlit \footnote{https://streamlit.io/}. We employ the MAWPS \cite{mawps} and ASDIV-a \cite{asdiva} datasets, which contains algebraic word problems of varying complexity curated from various websites. We filter out ungrammatical MWPs, yielding 1873 in MAWPS and 1844 train samples in ASDIV-a.
\begin{table}[hbt!]
\small
    \centering
    \begin{tabular}{c|c|c}
       \bf Dataset & \bf Method & \bf Accuracy (\%)  \\ \hline
        MAWPS \cite{mawps} & \textit{MWPRanker}& \textbf{83.76} \\
        & VectorSim \cite{huang-etal-2021-recall-learn} & 61.54 \\ \hline

                ASDIV-a \cite{asdiva} & \textit{MWPRanker}& \textbf{85.41} \\
        & VectorSim \cite{huang-etal-2021-recall-learn} & 42.71 \\ \hline
    \end{tabular}
    \caption{Performance Evaluation for MWP Similarity Detection}
    \label{tab:perf}
\end{table}
\subsection{Qualitative Analysis} 

We evaluated the tool with 19 graduate level users. The interface and an example are shown in a screenshot in Figure \ref{fig2}.  We can observe from recommended results that all MWPs have the same sequence of algebraic operations. The overall feedback in terms of ease of use and relevance of results was positive. Around \textbf{94\%} of the users found the tool easy to use and \textbf{84\%} found the tool to produce relevant results. About \textbf{15.8\%} of the users found the results to be relevant, with minor errors. Overall, all users rated that they would recommend the tool to the academicians. 

\subsection{Quantitative Analysis} 
For quantitative analysis, we collect 40 samples from a test set of MAWPS and ASDIV-a and use them as queries to retrieve top 3 questions from an MWP repository curated from the mentioned data sources. We present them to two independent researchers and asked them to annotate a recommendation as 1 if it is a similar (duplicate) MWP, else 0. We use the vector based semantic similarity based model proposed in \cite{huang-etal-2021-recall-learn} as a baseline. We observed a reasonable level of agreement between the annotators, with a Cohen's kappa of \textbf{0.629}. From Table \ref{tab:perf}, we observe that the proposed MWPRanker tool outperforms the semantic similarity based approach by a significant margin.

The quality of retrieval depends on the quality of algebraic expression generated by the neural expression generator. To generate better questions, complexity based attributes are necessary for more fine-grained retrieval, which are not currently supported by MWPRanker.
\section{Conclusion}
In this work, we propose a new task of retrieving Math Word Problems based on the similarity of the algebraic expression. We develop and deploy a tool for the same to aid in recommending more practice questions.  In the future, for ease of access to the tool, we plan to explore automated search completion and the usage of MWPRanker for automated problem-solving. This module can also be used to retrieve samples for In-Context learning in language models.
\bibliography{pkdd}
\bibliographystyle{splncs04}
\end{document}


\title{'John ate 5 apples' != 'John ate some apples': Self-Supervised Paraphrase Quality Detection for Algebraic Word Problems\\ \small Supplementary Material}
 \textbf{Supplementary Section}
\appendix

\begin{table*}[hbt!]
\small
\caption{An ablative analysis of each operator across all datasets. Here, each operator $f_i$ represents the results when we train after removing that operator. The numbers in bold represent the lowest scores for positive operators ($f_1, \dots f_4$) and negative operators ($f_5, \dots f_{10}$) each, thereby demonstrating the impact of that operator.}
\begin{tabular}{clrrrrrrrrp{0.6cm}}
\hline
\multicolumn{1}{l}{\multirow{2}{*}{\textbf{Dataset}}} & \multirow{2}{*}{\textbf{Op}} & \multicolumn{3}{c}{\textbf{Macro}} & \multicolumn{3}{c}{\textbf{Weighted}} & \multicolumn{1}{l}{\multirow{2}{*}{$\mu^+$}} & \multicolumn{1}{l}{\multirow{2}{*}{$\mu^-$}} & \multicolumn{1}{l}{\multirow{2}{*}{$\mu^s$}} \\ \cline{3-8}
\multicolumn{1}{l}{} &  & \multicolumn{1}{l}{P} & \multicolumn{1}{l}{R} & \multicolumn{1}{l}{F1} & \multicolumn{1}{l}{P} & \multicolumn{1}{l}{R} & \multicolumn{1}{l}{F1} & \multicolumn{1}{l}{} & \multicolumn{1}{l}{} & \multicolumn{1}{l}{} \\ \hline
\multirow{10}{*}{AquaRAT} & $f_1$ & 0.667 & 0.681 & 0.674 & 0.749 & 0.611 & 0.673 & 0.742 & 0.024 & 0.718 \\
 & $f_2$ & 0.682 & 0.694 & 0.688 & 0.769 & 0.616 & 0.684 & 0.813 & 0.047 & 0.766 \\
 & $f_3$ & 0.663 & 0.669 & \textbf{0.666} & 0.75 & 0.586 & \textbf{0.658} & 0.783 & 0.115 & \textbf{0.668} \\
 & $f_4$ & 0.679 & 0.686 & 0.682 & 0.768 & 0.602 & 0.675 & 0.827 & 0.084 & 0.744 \\
 & $f_5$ & 0.664 & 0.67 & 0.667 & 0.751 & 0.589 & 0.66 & 0.791 & 0.119 & 0.672 \\
 & $f_6$ & 0.667 & 0.669 & 0.668 & 0.756 & 0.582 & 0.658 & 0.813 & 0.138 & 0.675 \\
 & $f_7$ & 0.667 & 0.678 & 0.673 & 0.753 & 0.602 & 0.669 & 0.771 & 0.069 & 0.702 \\
 & $f_8$ & 0.671 & 0.679 & 0.675 & 0.758 & 0.598 & 0.668 & 0.796 & 0.09 & 0.706 \\
 & $f_9$ & 0.653 & 0.657 & \textbf{0.655} & 0.74 & 0.573 & \textbf{0.646} & 0.77 & 0.145 & \textbf{0.625} \\  
 & $f_{10}$ & 0.678 & 0.687 & 0.683 & 0.766 & 0.607 & 0.677 & 0.813 & 0.078 & 0.735 \\ \hline
\multirow{10}{*}{EM\_Math} & $f_1$ & 0.635 & 0.644 & 0.64 & 0.67 & 0.627 & 0.648 & 0.425 & -0.144 & 0.569 \\
 & $f_2$ & 0.648 & 0.651 & 0.65 & 0.689 & 0.613 & 0.649 & 0.598 & -0.005 & 0.603 \\
 & $f_3$ & 0.666 & 0.669 & 0.667 & 0.709 & 0.628 & 0.666 & 0.661 & -0.017 & 0.678 \\
 & $f_4$ & 0.638 & 0.635 & \textbf{0.636} & 0.681 & 0.588 & \textbf{0.631} & 0.633 & 0.096 & \textbf{0.538} \\
 & $f_5$ & 0.653 & 0.653 & 0.653 & 0.697 & 0.608 & 0.65 & 0.651 & 0.042 & 0.609 \\
 & $f_6$ & 0.657 & 0.652 & 0.655 & 0.703 & 0.602 & 0.648 & 0.696 & 0.088 & 0.608 \\
 & $f_7$ & 0.66 & 0.658 & 0.659 & 0.705 & 0.612 & 0.655 & 0.681 & 0.048 & 0.633 \\
 & $f_8$ & 0.646 & 0.648 & \textbf{0.647} & 0.688 & 0.608 & 0.646 & 0.606 & 0.018 & \textbf{0.588} \\
 & $f_9$ & 0.656 & 0.649 & 0.652 & 0.702 & 0.597 & \textbf{0.645} & 0.704 & 0.108 & 0.596 \\ 
 & $f_{10}$ & 0.672 & 0.674 & 0.673 & 0.716 & 0.632 & 0.671 & 0.677 & -0.012 & 0.689 \\ \hline
\multirow{10}{*}{SAWP} & $f_1$ & 0.618 & 0.627 & 0.623 & 0.688 & 0.572 & 0.625 & 0.572 & 0.06 & 0.512 \\
 & $f_2$ & 0.617 & 0.621 & \textbf{0.619} & 0.69 & 0.552 & \textbf{0.614} & 0.631 & 0.152 & \textbf{0.479} \\
 & $f_3$ & 0.624 & 0.63 & 0.627 & 0.697 & 0.565 & 0.624 & 0.632 & 0.114 & 0.518 \\
 & $f_4$ & 0.646 & 0.648 & 0.647 & 0.725 & 0.57 & 0.638 & 0.739 & 0.16 & 0.579 \\
 & $f_5$ & 0.648 & 0.644 & 0.646 & 0.729 & 0.56 & 0.633 & 0.777 & 0.205 & 0.572 \\
 & $f_6$ & 0.629 & 0.632 & 0.631 & 0.705 & 0.56 & 0.624 & 0.679 & 0.148 & 0.53 \\
 & $f_7$ & 0.63 & 0.637 & 0.634 & 0.704 & 0.572 & 0.632 & 0.649 & 0.096 & 0.553 \\
 & $f_8$ & 0.643 & 0.64 & 0.642 & 0.723 & 0.558 & 0.63 & 0.754 & 0.195 & 0.559 \\
 & $f_9$ & 0.626 & 0.622 & \textbf{0.624} & 0.705 & 0.538 & \textbf{0.61} & 0.723 & 0.245 & \textbf{0.478} \\ 
 & $f_{10}$ & 0.645 & 0.642 & 0.643 & 0.725 & 0.56 & 0.632 & 0.763 & 0.2 & 0.562 \\ \hline
\multirow{10}{*}{PAWP} & $f_1$ & 0.644 & 0.623 & \textbf{0.634} & 0.645 & 0.622 & \textbf{0.633} & 0.64 & 0.148 & \textbf{0.492} \\
 & $f_2$ & 0.68 & 0.636 & 0.658 & 0.681 & 0.635 & 0.657 & 0.769 & 0.226 & 0.543 \\
 & $f_3$ & 0.712 & 0.659 & 0.684 & 0.712 & 0.658 & 0.684 & 0.819 & 0.192 & 0.627 \\
 & $f_4$ & 0.714 & 0.654 & 0.683 & 0.714 & 0.652 & 0.682 & 0.847 & 0.224 & 0.623 \\
 & $f_5$ & 0.694 & 0.644 & 0.668 & 0.695 & 0.642 & 0.668 & 0.8 & 0.229 & 0.57 \\
 & $f_6$ & 0.705 & 0.646 & 0.674 & 0.706 & 0.645 & 0.674 & 0.829 & 0.244 & 0.585 \\
 & $f_7$ & 0.698 & 0.651 & 0.674 & 0.698 & 0.65 & 0.673 & 0.795 & 0.185 & 0.61 \\
 & $f_8$ & 0.702 & 0.629 & 0.663 & 0.702 & 0.628 & 0.663 & 0.859 & 0.343 & 0.516 \\
 & $f_9$ & 0.663 & 0.619 & 0.64 & 0.663 & 0.618 & 0.64 & 0.759 & 0.284 & 0.475 \\ 
 & $f_{10}$ & 0.655 & 0.587 &\textbf{0.619} & 0.656 & 0.585 & \textbf{0.618} & 0.845 & 0.493 & \textbf{0.352} \\ \hline
 \vspace{-2mm}
\end{tabular}

\label{table:ops}
\end{table*}
\begin{figure*}[hbt!]
   \begin{subfigure}[b]{\textwidth}
        \centering
        \includegraphics[width=0.5\linewidth]{images/emb_plots/AQUA/AQUA_methods_UDA_1.pdf}%
        \includegraphics[width=0.5\linewidth]{images/emb_plots/AQUA/AQUA_methods_UDA_2.pdf}
        \caption{UDA}
    \end{subfigure}
   \begin{subfigure}[b]{\textwidth}
        \centering
        \includegraphics[width=0.49\linewidth]{images/emb_plots/AQUA/AQUA_methods_SSMBA_1.pdf}
        \includegraphics[width=0.49\linewidth]{images/emb_plots/AQUA/AQUA_methods_SSMBA_2.pdf}
        \caption{SSMBA}
    \end{subfigure}
\caption{Further Embedding plots on AquaRAT}
\label{fig:plots2}
\vspace{-2mm}
\end{figure*}

\begin{figure*}[h!]
\begin{subfigure}{0.32\linewidth}
\hspace{-1.2cm}
\includegraphics[width=1.25\linewidth]{images/data_stats/AQUA/cm.png}
\caption{AquaRAT}
\end{subfigure}
\begin{subfigure}{0.32\linewidth}
\includegraphics[width=1.25\linewidth]{images/data_stats/EM/cm.png}
\caption{EM}
\end{subfigure}
\begin{subfigure}{0.32\linewidth}
\includegraphics[width=1.25\linewidth]{images/data_stats/SAWP/cm.png}
\caption{SAWP}
\end{subfigure}
\caption{Statistics for Test Operators for AquaRAT, EM and SAWP.}
\label{fig:test}
\end{figure*}

\begin{figure*}[h!]
\begin{subfigure}{0.32\linewidth}
\hspace{-2pt}
\includegraphics[width=1.5\linewidth]{images/confusion_matrices/AQUA/ParaQD (base).pdf}
\caption{ParaQD}
\end{subfigure}
\begin{subfigure}{0.32\linewidth}
\includegraphics[width=1.5\linewidth]{images/confusion_matrices/AQUA/SSMBA (w pl).pdf}
\caption{SSMBA with PL}
\end{subfigure}%
\hspace{1em}
\begin{subfigure}{0.32\linewidth}
\includegraphics[width=1.5\linewidth]{images/confusion_matrices/AQUA/UDA (w pl).pdf}
\caption{UDA w pl}
\end{subfigure}
\label{subfig}
\begin{subfigure}{0.32\linewidth}
\includegraphics[width=1.5\linewidth]{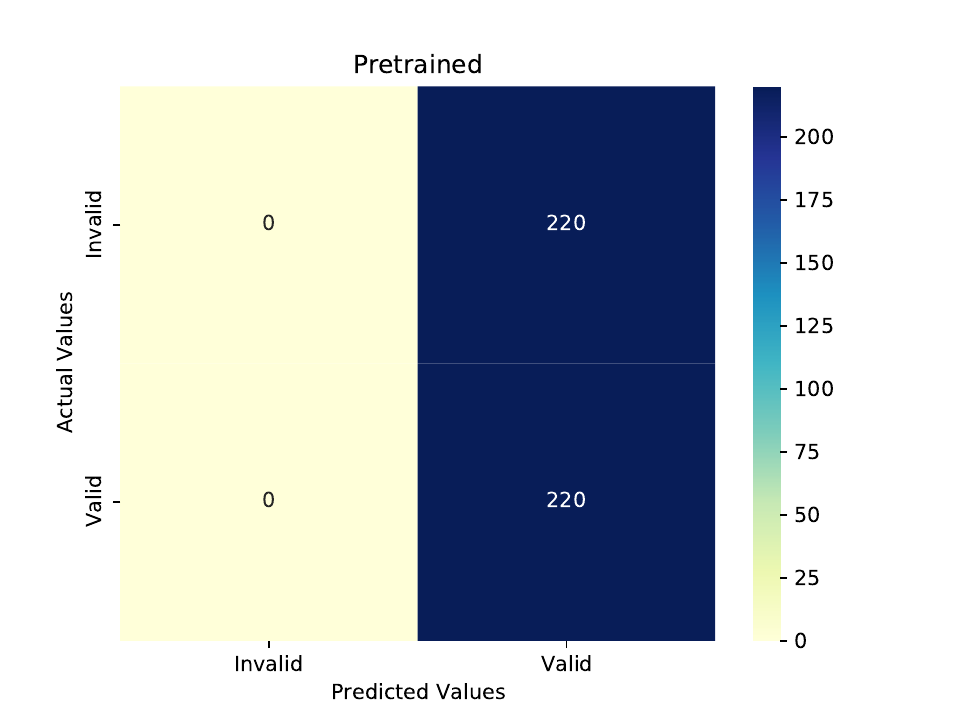}
\caption{Pretrained}
\end{subfigure}%
\begin{subfigure}{0.31\linewidth}
\includegraphics[width=1.5\linewidth]{images/confusion_matrices/AQUA/SSMBA (wo pl).pdf}
\caption{SSMBA}
\end{subfigure}
\begin{subfigure}{0.32\linewidth}
\includegraphics[width=1.5\linewidth]{images/confusion_matrices/AQUA/UDA (wo pl).pdf}
\caption{UDA}
\end{subfigure}%
\caption{Confusion matrices for all methods on AquaRAT}
\label{fig:cm}
\vspace{-2mm}
\end{figure*}

\begin{figure*}[hbt!]
\begin{subfigure}{0.32\linewidth}
\hspace{-2pt}
\includegraphics[width=1.5\linewidth]{images/confusion_matrices/EM/ParaQD (base).pdf}
\caption{ParaQD}
\end{subfigure}
\begin{subfigure}{0.32\linewidth}
\includegraphics[width=1.5\linewidth]{images/confusion_matrices/EM/SSMBA (with PL).pdf}
\caption{SSMBA with pl}
\end{subfigure}%
\hspace{1em}
\begin{subfigure}{0.32\linewidth}
\includegraphics[width=1.5\linewidth]{images/confusion_matrices/EM/UDA (with PL).pdf}
\caption{UDA with pl}
\end{subfigure}
\label{subfig}
\begin{subfigure}{0.32\linewidth}
\includegraphics[width=1.5\linewidth]{images/confusion_matrices/EM/Pretrained.pdf}
\caption{Pretrained}
\end{subfigure}%
\begin{subfigure}{0.31\linewidth}
\includegraphics[width=1.5\linewidth]{images/confusion_matrices/EM/SSMBA (wo pl).pdf}
\caption{SSMBA}
\end{subfigure}
\begin{subfigure}{0.32\linewidth}
\includegraphics[width=1.5\linewidth]{images/confusion_matrices/EM/UDA (wo pl).pdf}
\caption{UDA}
\end{subfigure}%
\hspace{1em}
\caption{Confusion matrices for all methods on EM\_Math.}
\label{fig:cm-EM}
\vspace{-2mm}
\end{figure*}

\begin{figure*}[hbt!]
\begin{subfigure}{0.32\linewidth}
\hspace{-2pt}
\includegraphics[width=1.5\linewidth]{images/confusion_matrices/SAWP/ParaQD (base).pdf}
\caption{ParaQD}
\end{subfigure}
\begin{subfigure}{0.32\linewidth}
\includegraphics[width=1.5\linewidth]{images/confusion_matrices/SAWP/SSMBA (w pl).pdf}
\caption{SSMBA with pl}
\end{subfigure}%
\hspace{1em}
\begin{subfigure}{0.32\linewidth}
\includegraphics[width=1.5\linewidth]{images/confusion_matrices/SAWP/UDA (w PL).pdf}
\caption{UDA with pl}
\end{subfigure}
\label{subfig}
\begin{subfigure}{0.32\linewidth}
\includegraphics[width=1.5\linewidth]{images/confusion_matrices/EM/Pretrained.pdf}
\caption{Pretrained}
\end{subfigure}%
\begin{subfigure}{0.31\linewidth}
\includegraphics[width=1.5\linewidth]{images/confusion_matrices/EM/SSMBA (wo pl).pdf}
\caption{SSMBA}
\end{subfigure}
\begin{subfigure}{0.32\linewidth}
\includegraphics[width=1.5\linewidth]{images/confusion_matrices/EM/UDA (wo PL).pdf}
\caption{UDA}
\end{subfigure}%
\hspace{1em}
\caption{Confusion matrices for all methods on SAWP.}
\label{fig:cm-SAWP}
\vspace{-2mm}
\end{figure*}

\begin{figure*}[hbt!]
\begin{subfigure}{0.32\linewidth}
\hspace{-2pt}
\includegraphics[width=1.5\linewidth]{images/confusion_matrices/PAWP/ParaQD (base).pdf}
\caption{ParaQD}
\end{subfigure}
\begin{subfigure}{0.32\linewidth}
\includegraphics[width=1.5\linewidth]{images/confusion_matrices/PAWP/SSMBA (w pl).pdf}
\caption{SSMBA with pl}
\end{subfigure}%
\hspace{1em}
\begin{subfigure}{0.32\linewidth}
\includegraphics[width=1.5\linewidth]{images/confusion_matrices/PAWP/UDA (w pl).pdf}
\caption{UDA with pl}
\end{subfigure}
\label{subfig}
\begin{subfigure}{0.32\linewidth}
\includegraphics[width=1.5\linewidth]{images/confusion_matrices/PAWP/Pretrained.pdf}
\caption{Pretrained}
\end{subfigure}%
\begin{subfigure}{0.31\linewidth}
\includegraphics[width=1.5\linewidth]{images/confusion_matrices/PAWP/SSMBA (wo pl).pdf}
\caption{SSMBA}
\end{subfigure}
\begin{subfigure}{0.32\linewidth}
\includegraphics[width=1.5\linewidth]{images/confusion_matrices/PAWP/UDA (wo pl).pdf}
\caption{UDA}
\end{subfigure}%
\hspace{1em}
\caption{Confusion matrices for all methods on PAWP}
\label{fig:cm-PAWP}
\vspace{-2mm}
\end{figure*}

\section{Augmentation: Further Details}
\label{appendix:augs}

\subsection{$f_1:$ Backtranslation}
We used WMT'19 FSMT \cite{ng2019facebook} \textit{en-de} and \textit{de-en} translation models, with language $A$ being $English$ and $B$ being $German$.
We used diverse beam search \cite{vijayakumar2018diverse} for decoding in the reverse translation step to introduce diversity and chose the candidate paraphrase with the maximum Levenshtein distance. This ensures that the model learns to give a high score for diverse paraphrases that retain critical information.

\subsection{$f_5$: Most Important Phrase Deletion}
To model $\Psi$, we use TopicRank \cite{bougouin-etal-2013-topicrank}. The methodology to select the most critical phrase is inspired by TSDAE \cite{wang2021tsdae}.

\subsection{$f_b$: Corrupted Sentence Reconstruction}
When corrupting the input sentence, we preserve the numbers, units and the last three tokens. This is done because if we corrupt the numbers or the units, the model cannot accurately reconstruct them and will replace them with random numbers and units. We preserve the last three tokens because corrupting them might lead the model to change the question as the last three tokens in a word problem are generally indicative of the question.


\section{Baselines: Pseudo Labelling}
\label{appendix:pl}
We use the same pretrained encoder (MiniLM) to first pseudo-label the samples (without being trained) and then train it using the pseudo-labelled samples. More formally, given an input $\Q$ and a paraphrase $\Q'$, we use the encoder to determine whether $\Q'$ is a positive or negative paraphrase of $\Q$ as follows:
  \begin{align*}
    &\rho_i, \zeta_i = ENC(\Q), ENC(\Q') \\
    &\lambda(\Q, \Q')=
    \begin{cases}
      1 &if\;cossim(\rho_i, \zeta_i) > \iota \\
      0 &if\;cossim(\rho_i, \zeta_i) \leq \iota 
     \end{cases}
  \end{align*}
where $\iota$ is the threshold for the cosine similarity, which we set to 0.8.

We observe that on AquaRAT, the performance decreases for both UDA and SSMBA due to pseudo labelling, while it increases on EM\_Math dataset. This can be due to the much higher percentage of pseudo-labelled negative samples for EM\_Math as shown in Table \ref{table:pl}, thus providing more information about detecting invalid paraphrases as seen in Figure \ref{fig:cm-EM}.

\section{Embedding Plots}
\label{appendix:plots}
We plotted the embeddings across triplets in the test set to observe the separation margin. The colour represents a triplet while the symbol represents which component of the triplet it is (anchor, positive, negative). The left embedding plot is for 7 randomly chosen triplets in the data (to closely visualize the distances), while the one on the right is for all triplets (93).

\begin{table*}[htb!]
\caption{Effect of the Loss Function for ParaQD (on AquaRAT)}
\begin{tabular}{crrrrrrllp{0.6cm}}
\hline
\multirow{2}{*}{\textbf{Loss}} & \multicolumn{3}{c}{\textbf{Macro}} & \multicolumn{3}{c}{\textbf{Weighted}} & \multirow{2}{*}{\textbf{$\mu^+$}} & \multirow{2}{*}{\textbf{$\mu^-$}} & \multirow{2}{*}{\textbf{$\mu^s$}} \\ \cline{2-7}
 & \multicolumn{1}{l}{\textbf{P}} & \multicolumn{1}{l}{\textbf{R}} & \multicolumn{1}{l}{\textbf{F1}} & \multicolumn{1}{l}{\textbf{P}} & \multicolumn{1}{l}{\textbf{R}} & \multicolumn{1}{l}{\textbf{F1}} &  &  &  \\ \hline
Triplet & 0.678 & 0.695 & 0.687 & 0.762 & 0.625 & 0.687 & \multicolumn{1}{r}{0.77} & \multicolumn{1}{r}{-0.01} & \multicolumn{1}{r}{\textbf{0.78}} \\ \hline
MultipleNegativeRankingLoss & 0.708 & 0.716 & \textbf{0.712} & 0.801 & 0.627 & \textbf{0.704} & \multicolumn{1}{r}{0.89} & \multicolumn{1}{r}{0.474} & \multicolumn{1}{r}{0.416} \\ \hline
\end{tabular}
\label{table:loss}
\end{table*}


\begin{table*}[htb!]
\caption{An ablative analysis of all methods for different seeds on AquaRAT.}
\begin{tabular}{llrrrrrrrrp{0.6cm}}
\hline
\multirow{2}{*}{\textbf{Seed}} &
  \multirow{2}{*}{\textbf{Method}} &
  \multicolumn{3}{c}{\textbf{Macro}} &
  \multicolumn{3}{c}{\textbf{Weighted}} &
  \multicolumn{1}{l}{\multirow{2}{*}{\textbf{$\mu^+$}}} &
  \multicolumn{1}{l}{\multirow{2}{*}{\textbf{$\mu^-$}}} &
  \multicolumn{1}{l}{\multirow{2}{*}{\textbf{$\mu^s$}}} \\ \cline{3-8}
 &
   &
  \multicolumn{1}{l}{\textbf{P}} &
  \multicolumn{1}{l}{\textbf{R}} &
  \multicolumn{1}{l}{\textbf{F1}} &
  \multicolumn{1}{l}{\textbf{P}} &
  \multicolumn{1}{l}{\textbf{R}} &
  \multicolumn{1}{l}{\textbf{F1}} &
  \multicolumn{1}{l}{} &
  \multicolumn{1}{l}{} &
  \multicolumn{1}{l}{} \\ \hline
\multicolumn{1}{r}{\multirow{3}{*}{3407}} & ParaQD & 0.678 & 0.695 & \textbf{0.687} & 0.762 & 0.625 & \textbf{0.687} & 0.77  & -0.01 & \textbf{0.78}  \\
\multicolumn{1}{r}{}                      & UDA    & 0.661 & 0.512 & 0.577          & 0.786 & 0.332 & 0.467          & 0.995 & 0.966 & 0.029          \\  
\multicolumn{1}{r}{}                      & SSMBA  & 0.645 & 0.554 & 0.596          & 0.757 & 0.395 & 0.52           & 0.965 & 0.829 & 0.137          \\ \hline
\multirow{3}{*}{Seed Search}              & ParaQD & 0.684 & 0.694 & \textbf{0.689} & 0.772 & 0.614 & \textbf{0.684} & 0.828 & 0.055 & \textbf{0.772} \\
                                          & UDA    & 0.659 & 0.503 & 0.571          & 0.784 & 0.32  & 0.455          & 0.998 & 0.985 & 0.013          \\  
                                          & SSMBA  & 0.634 & 0.552 & 0.59           & 0.742 & 0.395 & 0.516          & 0.957 & 0.833 & 0.124          \\ \hline
\end{tabular}

\label{table:seed}
\end{table*}

\section{Loss Functions}
\label{appendix:loss}
Multiple Negatives Ranking Loss \footnotemark is a loss function, which, for anchor $a_i$ in the triplet $(a_i, p_i, n_i)$ considers $p_i$ as a positive sample and all $p_j$ and $n_k$ in the batch (such that $j!=i$) as negatives. It works by maximizing the log-likelihood of the softmax scores. The equation is similar to the one in \cite{henderson2017efficient}. 

\note{\url{https://www.sbert.net/docs/package\_reference/losses.html\#multiplenegativesrankingloss}}

\section{Performance on test set with training operators}
\label{appendix:trainops}
We also evaluated our method by generating the test set of AquaRAT using training operators. The results are present in Table \ref{table:trainops}. ParaQD outperforms the baselines by a significant margin across all metrics. However, this test set is not representative of real data distribution as it is suited for our method. Thus, we report the results only for the sake of completion and they should not be taken as representative.

\begin{table*}[htb!]
\caption{Performance of all methods on the test set of AquaRAT created using train operators.}
\begin{tabular}{lrrrrrrrrp{0.6cm}}
\hline
\multirow{2}{*}{\textbf{Method}} &
  \multicolumn{3}{c}{\textbf{Macro}} &
  \multicolumn{3}{c}{\textbf{Weighted}} &
  \multicolumn{1}{l}{\multirow{2}{*}{\textbf{$\mu^+$}}} &
  \multicolumn{1}{l}{\multirow{2}{*}{\textbf{$\mu^-$}}} &
  \multicolumn{1}{l}{\multirow{2}{*}{\textbf{$\mu^s$}}} \\ \cline{2-7}
 &
  \multicolumn{1}{l}{\textbf{P}} &
  \multicolumn{1}{l}{\textbf{R}} &
  \multicolumn{1}{l}{\textbf{F1}} &
  \multicolumn{1}{l}{\textbf{P}} &
  \multicolumn{1}{l}{\textbf{R}} &
  \multicolumn{1}{l}{\textbf{F1}} &
  \multicolumn{1}{l}{} &
  \multicolumn{1}{l}{} &
  \multicolumn{1}{l}{} \\ \hline
Pretrained    & 0.25  & 0.5   & 0.333          & 0.25  & 0.5            & 0.333 & 0.966 & 0.92   & 0.046          \\ 
UDA           & 0.626 & 0.505 & 0.559          & 0.626 & 0.505          & 0.559 & 0.995 & 0.973  & 0.022          \\ 
UDA (w pl)    & 0.681 & 0.511 & 0.584          & 0.681 & 0.511          & 0.584 & 0.99  & 0.965  & 0.025          \\ 
SSMBA         & 0.667 & 0.532 & 0.592          & 0.667 & 0.532          & 0.592 & 0.965 & 0.871  & 0.094          \\ 
SSMBA (w pl)  & 0.705 & 0.518 & 0.597          & 0.705 & 0.518          & 0.597 & 0.987 & 0.927  & 0.06           \\ 
ParaQD (ours) & 0.903 & 0.895 & \textbf{0.899} & 0.903 & 0.895 & \textbf{0.899} & 0.927 & -0.656 & \textbf{1.583} \\ \hline
\end{tabular}

\label{table:trainops}
\end{table*}

\begin{table*}[htb!]
\caption{An ablative analysis of all methods for 3 different encoders on AquaRAT. We observe that regardless of the encoder used, we outperform the baselines on all metrics.}
\hspace{-1cm}
\begin{tabular}{llrrrrrrrrp{0.6cm}}
\hline
\multirow{2}{*}{\textbf{Encoder}} &
  \multirow{2}{*}{\textbf{Method}} &
  \multicolumn{3}{c}{\textbf{Macro}} &
  \multicolumn{3}{c}{\textbf{Weighted}} &
  \multicolumn{1}{l}{\multirow{2}{*}{\textbf{$\mu^+$}}} &
  \multicolumn{1}{l}{\multirow{2}{*}{\textbf{$\mu^-$}}} &
  \multicolumn{1}{l}{\multirow{2}{*}{\textbf{$\mu^s$}}} \\ \cline{3-8}
 &
   &
  \multicolumn{1}{l}{\textbf{P}} &
  \multicolumn{1}{l}{\textbf{R}} &
  \multicolumn{1}{l}{\textbf{F1}} &
  \multicolumn{1}{l}{\textbf{P}} &
  \multicolumn{1}{l}{\textbf{R}} &
  \multicolumn{1}{l}{\textbf{F1}} &
  \multicolumn{1}{l}{} &
  \multicolumn{1}{l}{} &
  \multicolumn{1}{l}{} \\ \hline
\multirow{3}{*}{\textbf{all-minilm-L12-v1 (base)}} & ParaQD & 0.678 & 0.695 & \textbf{0.687} & 0.762 & 0.625 & \textbf{0.687} & 0.77  & -0.01 & \textbf{0.78}  \\
                                                   & UDA    & 0.661 & 0.512 & 0.577          & 0.786 & 0.332 & 0.467          & 0.995 & 0.966 & 0.029          \\ 
                                                   & SSMBA  & 0.645 & 0.554 & 0.596          & 0.757 & 0.395 & 0.52           & 0.965 & 0.829 & 0.137          \\ \hline
\multirow{3}{*}{\textbf{MPNet}}                    & ParaQD & 0.703 & 0.726 & \textbf{0.714} & 0.785 & 0.659 & \textbf{0.717} & 0.858 & 0.201 & \textbf{0.656} \\
                                                   & UDA    & 0.659 & 0.503 & 0.571          & 0.784 & 0.32  & 0.455          & 0.99  & 0.953 & 0.037          \\ 
                                                   & SSMBA  & 0.66  & 0.508 & 0.574          & 0.785 & 0.327 & 0.462          & 0.985 & 0.94  & 0.045          \\ \hline
\multirow{3}{*}{\textbf{all-minilm-L6-v2}}         & ParaQD & 0.671 & 0.679 & \textbf{0.675} & 0.758 & 0.598 & \textbf{0.668} & 0.799 & 0.083 & \textbf{0.716} \\
                                                   & UDA    & 0.659 & 0.503 & 0.571          & 0.784 & 0.32  & 0.455          & 0.994 & 0.979 & 0.015          \\ 
                                                   & SSMBA  & 0.661 & 0.513 & 0.578          & 0.786 & 0.334 & 0.469          & 0.992 & 0.941 & 0.051          \\ \hline
\end{tabular}

\label{table:encoder}
\end{table*}

\begin{table*}[htb!]
\centering
\caption{Pseudo Labelling Statistics. Positive\% represents the percentage of total samples pseudo-labelled as positive, while Negative\% represents the percentage of total samples pseudo-labelled as negatives.}
\begin{tabular}{llrp{0.6cm}}
\hline
Dataset                   & Method       & \multicolumn{1}{l}{Positive \%} & \multicolumn{1}{l}{Negative \%} \\ \hline
\multirow{2}{*}{AquaRAT}  & UDA (w pl)   & 87.29                           & 12.71                           \\
                          & SSMBA (w pl) & 75.69                           & 24.31                           \\ \hline
\multirow{2}{*}{EM\_Math} & UDA (w pl)   & 72.16                           & 27.84                           \\ \cline{2-4} 
                          & SSMBA (w pl) & 55.09                           & 44.91                           \\ \hline
\end{tabular}

\label{table:pl}
\end{table*}

\section{Encoder Ablations and Seed Optimization}
\label{appendix:encoder}
The results of varying the encoder are shown in Table \ref{table:encoder}. We vary the encoder and experiment with MiniLM (12 layers), MiniLM (6 layers) and MPNet \cite{song2020mpnet}. We choose the encoders considering different metrics such as average performance on semantic search and encoding speed ($\# \ of sentences/sec$)\footnote{The metrics are obtained from \url{https://www.sbert.net/docs/pretrained\_models.html}}. For instance MPNet can encode about \textit{2500} sentences/sec whereas MiniLM (12 layers) (all-minilm-L12-v1) can encode about 7500 sentences/sec and MiniLM (6 layers) (all-minilm-L6-v2) can encode about 14200 sentences/sec. Also the average performance on semantic search benchmarks is of order $all-minilm-L12-v1$ $>$ $all-minilm-L6-v2$ $>$ MPNet. In our setting, from table \ref{table:encoder} we can observe that MiniLM (12 layers) surpasses the other encoders as measured by the separation metric ($\mu^s$). However, when we observe Macro and weighted F1 scores MPNet surpasses the other two encoders. Since we are more concerned about how positive and negative paraphrases are separated in the vector space we choose MiniLM (12 layers) (all-minilm-L12-v1) for all our main experiments as shown in Table 1. We can also observe that the proposed method (ParaQD) outperforms all other baselines. This demonstrates the robustness of the proposed augmentation method and shows the performance gain when compared to other baselines is invariant to changes in encoders.
We also vary the seed values to check for robustness of our method. We compare the random seed value (3407) with the seed optimization method proposed in the Augmented SBERT paper \cite{thakur2021augmented}. For seed optimization, we search for the best seed in range [0-4] as recommended in the original work by training 20\% of the data and comparing the results on the validation set. We then select the best performing seed and train using that particular seed. The results of the experiments are shown in Table \ref{table:seed}. We observe that the best seed obtained through seed optimization and the seed value of 3407 nearly yield similar performance. We also observe that the proposed method outperforms the baselines demonstrating the robustness of the proposed method to seed randomization.

\bibliography{pkdd}
\bibliographystyle{splncs04}